\documentclass[aps,prl,twocolumn,superscriptaddress,floatfix,10pt]{revtex4-2}
\usepackage{amsmath}
\usepackage{amssymb}
\usepackage{graphicx}
\usepackage{booktabs}
\usepackage{hyperref}
\usepackage{color}
\usepackage{xcolor}

\begin{document}

\title{Universal Properties of Near-Threshold Single-Neutron Resonances}

\author{Myungkuk Kim}
\affiliation{Center for Exotic Nuclear Studies, Institute for Basic Science, Daejeon 34126, Korea}

\author{Young-Ho Song}
\affiliation{Institute for Rare Isotope Sciences, Institute for Basic Science, Daejeon 34000, Korea}

\author{Hans-Werner Hammer}
\affiliation{Technische Universit\"at Darmstadt, Department of Physics, 64289 Darmstadt, Germany}
\affiliation{ExtreMe Matter Institute EMMI and Helmholtz Forschungsakademie Hessen f\"ur FAIR (HFHF), GSI Helmholtzzentrum f\"ur Schwerionenforschung GmbH, 64291 Darmstadt, Germany}

\author{Youngman Kim}
\affiliation{Center for Exotic Nuclear Studies, Institute for Basic Science, Daejeon 34126, Korea}

\author{Dean Lee}
\affiliation{Facility for Rare Isotope Beams and Department of Physics and Astronomy, Michigan State University, East Lansing, MI 48824, USA}

\author{Yuan-Zhuo Ma}
\affiliation{Facility for Rare Isotope Beams and Department of Physics and Astronomy, Michigan 
State University, East Lansing, MI 48824, USA}

\date{\today}

\begin{abstract}
We establish universal width predictions for near-threshold single-neutron resonances in $L > 0$ partial waves. Our results go beyond Wigner’s well-known scaling behavior of cross sections near threshold. We show that the finite square-well potential exhibits discrete scale invariance at zero energy.  From this fact, we derive an analytic baseline for the resonance width that depends only on geometry, angular momentum, and resonance energy, and not on internal short-distance nuclear details or radial excitation. This is a nontrivial property that is unique to the finite square-well potential and does not occur for other potentials.  Application to observed $p$-wave and $d$-wave resonances demonstrates that the square-well result provides a robust baseline. We show that discrete scale invariance erases radial-node information in the sharp-boundary limit, but realistic Woods-Saxon diffuseness breaks this invariance, suppressing the reduced width by a factor sensitive to the internal radial excitation. These results provide a simple geometric benchmark for identifying when observed neutron resonances are controlled by universal threshold physics and when they exhibit systematic deviations driven by structure-dependent effects.
\end{abstract}

\maketitle

What, if anything, is universal about the physics of single-neutron resonances close to threshold? While the wave function of any resonance inherently extends to infinity, as the energy approaches zero, the diverging wavelength causes the decay dynamics to be dictated by the external geometry of the potential barrier \cite{Jensen:2004zz}. In this near-threshold regime, the short-distance details of the strong nuclear interaction decouple from the long-range asymptotic behavior. This decoupling suggests that observables such as the decay width might be dictated by universal geometric constraints rather than by detailed internal structure \cite{Wigner:1948zz,Braaten:2004rn}.

However, rigorously isolating these universal kinematic features from the specific dynamics of the continuum is a theoretical challenge. Orbital angular momentum plays a key role in the manifestation of any universal threshold physics. Because neutral $s$-wave interactions lack a confining 
barrier, near-threshold $L=0$ poles generally appear as bound or virtual states rather than shape resonances. However, in the case of short-range $s$-wave and long-range Coulomb interactions, universal resonances are known to exist \cite{Higa:2008dn,Schmickler:2019ewl,Mochizuki:2024dbf}. Since we focus on systems with one charged particle at most, we restrict our universal width analysis to partial waves with nonzero orbital angular momentum, $L>0$, where the centrifugal barrier provides the spatial trapping required to form a shape resonance.

In this Letter, we demonstrate and exploit the discrete scale invariance of zero-energy scattering for a finite square-well potential to establish predictions for the decay widths of $L > 0$ single-neutron resonances. By defining an effective channel boundary, we project a phenomenological baseline across the nuclear chart based on the analytic properties of the square-well potential \cite{nussenzveig1959poles}. We demonstrate that this baseline is independent of the radial excitation mode. While sufficient for low angular momentum ($L=1$) states, realistic Woods-Saxon (WS) geometries induce a partial-wave-dependent suppression mechanism that breaks this universality, providing a diagnostic tool for internal nuclear structure.

For a finite-range interaction, the resonance width near threshold is constrained by the low-energy scattering parameters. We use units with $\hbar=c=1$ unless otherwise stated. The scattering phase shift $\delta_L(k)$ for relative momentum $k = \sqrt{2\mu E}$ is parameterized by the effective range expansion, 
\begin{equation}
    k^{2L+1} \cot \delta_L(k) = -\frac{1}{a_L} + \frac{1}{2} r_L k^2 + \dots,
\end{equation}
where $a_L$ is the scattering length and $r_L$ is the effective range parameter. A narrow resonance at energy $E_r$ occurs when $\cot \delta_L(k_r) = 0$. The analytic continuation to the $S$-matrix pole at $E_{pole} \approx E_r - i\Gamma_L/2$ dictates that the decay width is governed by the effective range: $\Gamma_L \approx 2 k_r^{2L+1} / (\mu |r_L|)$. Predicting the near-threshold width reduces to determining the effective range.

The analytic square-well relation is universal for a finite-range interaction with range $R$. Based on causality constraints and generalized Wigner bounds \cite{Wigner:1955zz, Phillips:1996ae,Hammer:2009zh,Hammer:2010fw}, the effective range at the zero-energy threshold ($|a_L| \to \infty$) is limited by the exterior scattering solutions. We then have the rigorous upper limit $r_L \le b_L(R)$, where
\begin{equation}
    b_L(R) = -\frac{2\Gamma(L-1/2)\Gamma(L+1/2)}{\pi} \left(\frac{R}{2}\right)^{1-2L}.
\end{equation}

As detailed in the Supplemental Material, for a square well tuned to threshold, the internal normalization of the wave function links the effective range to this exterior geometry, yielding the scale-invariant relation:
\begin{equation}
    r_L = -\left(L + \frac{1}{2}\right) |b_L(R)|.
\label{eq:sharpened_r}
\end{equation}
We note that the radial excitation number does not appear in this formula.  Substituting Eq.~(\ref{eq:sharpened_r}) into the pole width relation yields a closed-form formula for the resonance width:
\begin{equation}
    \Gamma_L = \frac{\pi (2\mu E_r)^{L+1/2} \left(\frac{R}{2}\right)^{2L-1}}{\mu (L + 1/2) \Gamma(L-1/2)\Gamma(L+1/2)},
    \label{eq:master_formula}
\end{equation}
where $\mu$ is the reduced mass of the core-valence system. Factors of $\hbar c$ are restored in numerical evaluations. 

The theoretical utility of Eq.~(\ref{eq:master_formula}) lies in its discrete scale invariance. For a square well at threshold ($E \to 0$), the continuity of the wave function requires the interior momentum $K_0 = \sqrt{2\mu V_0}$ to satisfy $K_0 R = \chi_{n,L-1}$, where $\chi_{n,L-1}$ is the $(n+1)$-th root of the spherical Bessel function $j_{L-1}(z)$ and $n$ corresponds to the radial excitation node. As explicitly proven in the Supplemental Material, this constraint allows the internal momentum to be scaled by a discrete transformation factor $\lambda = \chi_{n',L-1} / \chi_{n,L-1}$. This exact discrete mapping maps the internal wave function of an $n$-node state onto an $n'$-node state without altering the external boundary matching. Because this discrete symmetry physically decouples the internal radial excitation from the external decay dynamics, the potential depth $V_0$ completely vanishes from the observable width. Consequently, the square-well width prediction relies solely on the external geometry and remains independent of $n$. For any potential with a varying internal depth, the local momentum remains inherently coupled to the macroscopic radius, fundamentally breaking this discrete mapping and restoring the width's dependence on the internal radial excitation.  

This discrete scale invariance for the square-well potential at threshold is similar to the discrete scale symmetry appearing in the Efimov effect and related phenomena \cite{Efimov:1970zz,Efimov:1971zz,kraemer2006evidence,braaten2007efimov,efimov2009giant,hammer2010efimov,wang2012origin,nishida2011liberating, nishida2012weakly,kunitski2015observation,naidon2017efimov,kievsky2021efimov}. In the Efimov effect, excited-state wave functions are also mapped onto each other by discrete scale transformations. A notable difference is that the discrete scale factor for the Efimov effect is a universal constant for all consecutive excited states, whereas the scale factor in our framework depends explicitly on the specific radial excitation numbers involved. When Efimov states cross the three-body threshold at negative scattering length, they turn into $s$-wave resonances whose energy and width follows a universal scaling law \cite{Bringas:2004zz, Deltuva:2020sdd, Dietz:2021haj}. However, as they cross the threshold, they broaden rapidly and dissociate into the scattering continuum.

\begin{table*}[t]
\centering
\begin{tabular}{@{}lcccccccc@{}}
\toprule
Isotope & $J^\pi$ & $L$ & $A_{\text{core}}$ & $R_{\text{eff}}$ (fm) & $E_r^{\text{exp}}$ (MeV) & $\Gamma^{\text{exp}}$ (MeV) & $\Gamma^{\text{pred}}_{\text{SW}}$ (MeV) & Ref. \\ \midrule
$^{5}$He  & $3/2^-$ & 1 & 4  & 2.957 & 0.735(30) & 0.719(30) & 0.488(30) & \cite{Tilley:2002vg} \\
$^{7}$He  & $3/2^-$ & 1 & 6  & 3.219 & 0.430(3)  & 0.182(5)  & 0.246(03) & \cite{CAO201246} \\
$^{9}$He  & $1/2^-$ & 1 & 8  & 3.428 & 1.28(1)   & 0.82(4)   & 1.371(16) & \cite{Sun:2026ntq} \\
$^{13}$Be & $1/2^-$ & 1 & 12 & 3.759 & 0.51(1)   & 0.45(3)   & 0.385(11) & \cite{Kondo:2010, Corsi:2019bto} \\
$^{28}$F  & $1^-$   & 1 & 27 & 4.571 & 0.199(6)  & 0.18(4)   & 0.117(05)  & \cite{PhysRevLett.124.152502, Delion:2023vse} \\ \midrule
$^{11}$Be & $5/2^+$ & 2 & 10 & 3.605 & 1.276(1)  & 0.10(1)   & 0.2112(4)  & \cite{Kelley:2012zz} \\
$^{15}$Be & $5/2^+$ & 2 & 14 & 3.897 & 1.80(10)  & 0.575(200) & 0.656(91) & \cite{Snyder:2013} \\
$^{25}$O  & $3/2^+$ & 2 & 24 & 4.439 & 0.749(10) & 0.088(6)  & 0.113(04)  & \cite{Kondo:2016vzi, Hoffman:2008szc} \\
\bottomrule
\end{tabular}
\caption{Comparison of experimental parameters to square-well predictions using Eq.~(\ref{eq:master_formula}) with $R = R_{\text{eff}} = 1.14(A_{\text{core}}^{1/3} + 1)$ fm and $\mu = \frac{A_{\text{core}}}{A_{\text{core}}+1} m_n$. For both $L=1$ and $L=2$ partial waves, the geometric baseline successfully predicts the proper physical scale of the widths within a factor of two. Experimental uncertainties are denoted in parentheses.}
\label{tab:results}
\end{table*}

Applying our universal width formula to nuclei requires an estimate of the effective finite well radius, $R_{\text{eff}}$, and the reduced mass, $\mu = \frac{A_{\text{core}}}{A_{\text{core}}+1} m_n$. For this, we define the channel radius as $R_{\text{eff}} = R_0 (A_{\text{core}}^{1/3} + 1)$, where $R_0$ is derived from the standard liquid-drop nuclear saturation density $\rho_0 = 0.16 \text{ fm}^{-3}$. Using the relation $\frac{4}{3}\pi R_0^3 = \rho_0^{-1}$, we obtain a radius parameter of $R_0 \approx 1.14 \text{ fm}$. Table~\ref{tab:results} compares the square-well prediction ($\Gamma^{\text{pred}}_{\text{SW}}$) to established single-neutron resonances with reported near-threshold energies and widths for which a dominant $L>0$ assignment is available. We emphasize that this theoretical estimate strictly applies to resonances that are predominantly single-neutron in character. Consequently, ambiguous spin-parity assignments and strongly fragmented states are excluded from the benchmark set.

For both $p$-wave ($L=1$) and $d$-wave ($L=2$) states across the light mass region, the scale-invariant square-well formula successfully captures the proper physical scale. While the data is not perfectly reproduced—reflecting expected spectroscopic fragmentation and continuum coupling—the central result is that the geometric baseline reliably estimates the experimental widths within a factor of two across all cases. 

Deviations from this idealized sharp-boundary approximation point to structural effects, such as the subtle role of the diffuse nuclear surface. The observable decay width is determined by the standard R-matrix factorization \cite{Lane:1958vxu}, $\Gamma_L = 2 P_L \gamma^2$. Here, the reduced width ($\gamma^2$) reflects the probability amplitude of the neutron at the nuclear surface, while the penetrability ($P_L$) dictates the probability of tunneling through the external barrier \cite{Gurvitz:1986uv}.

A realistic Woods-Saxon (WS) potential has a soft, diffuse edge rather than a rigid boundary. This diffuse surface introduces two competing geometric effects. First, the internal wave function spreads into the soft edge, diluting the probability density at the boundary and thereby suppressing the reduced width ($\gamma^2$). Second, this same diffuse tail alters the location of the inner classical turning point and thins the external potential barrier, enhancing the penetrability ($P_L$).

The balance of this competition depends sensitively on angular momentum. For $p$-wave ($L=1$) states, the centrifugal barrier is relatively weak. The neutron resides predominantly within the uniform nuclear core, causing the $\gamma^2$ suppression and $P_L$ enhancement to roughly cancel out. Consequently, the scale-invariant square-well baseline remains accurate. 

Conversely, for $d$-wave ($L=2$) states, the stronger $L(L+1)/r^2$ centrifugal repulsion physically pushes the inner classical turning point outward. This forces the internal wave function to increasingly sample the diffuse surface region. Because the $L=2$ state is squeezed into this soft boundary, the dilution of the internal probability density overpowers the enhanced penetrability, providing a geometric mechanism for decay width suppression. As derived in the Supplemental Material, evaluating the difference in WKB normalization integrals over the diffuse surface yields an estimate for the WS reduced width scaled by:
\begin{equation}
    \gamma^2_{\text{WS}} \approx \gamma^2_{\text{SW}} \left\{ 1 + \frac{2a \chi_{n, L-1}}{R_{\text{eff}}\sqrt{L(L+1)}} \right\}^{-1}.
\label{eq:ws_suppression}
\end{equation}

Equation~(\ref{eq:ws_suppression}) demonstrates that WS suppression is sensitive to the internal radial structure. Higher radial nodes ($n$) correspond to larger internal momenta ($\chi_{n,L-1}$) and are more sensitive to the diffuse surface. Discrete scale invariance erases radial-node information in the sharp-boundary limit, but diffuseness breaks this invariance and makes the width sensitive to the internal radial excitation.

Evaluating Eq.~(\ref{eq:ws_suppression}) with standard nuclear parameters ($a \approx 0.65$ fm) provides a semi-quantitative estimate of this structural suppression. For a $d$-wave resonance ($L=2$) assuming the lowest radial mode ($n=0$, $\chi_{0,1} \approx 4.49$), the WS diffuseness yields a geometric suppression factor of roughly $1.8$. Applying this estimate provides a geometric mechanism for width suppression relative to the sharp-boundary limit. However, given that the baseline geometric estimate already works within a factor of two for the data, we view this WS diffuseness as one of several competing structural phenomena—alongside core excitation and spectroscopic fragmentation—that drive the observed scatter around the universal baseline.

While this geometric framework provides a robust baseline for light nuclei, extending this analysis to heavier systems faces simultaneous theoretical and experimental challenges. Theoretically, as the core mass increases, the level density explodes and the single-particle strength becomes heavily fragmented into complex compound-nuclear resonances. Experimentally, the neutron dripline for heavier masses has historically been exceedingly difficult to reach. The resulting low production cross-sections yielded limited statistics, severely restricting the resolution required to isolate narrow, near-threshold shape resonances from the continuum background. Consequently, dripline systems in the light mass region have served as the primary laboratory for isolating these universal kinematic features. However, the capabilities of next-generation radioactive ion beam facilities will systematically push the observable dripline into these heavier mass regions. As high-intensity beams make these exotic isotopes accessible with sufficient statistics, the geometric framework established here will provide a crucial diagnostic baseline for separating universal threshold physics from emergent many-body complexity.

\begin{figure*}[t]
    \centering
    \includegraphics[width=14cm]{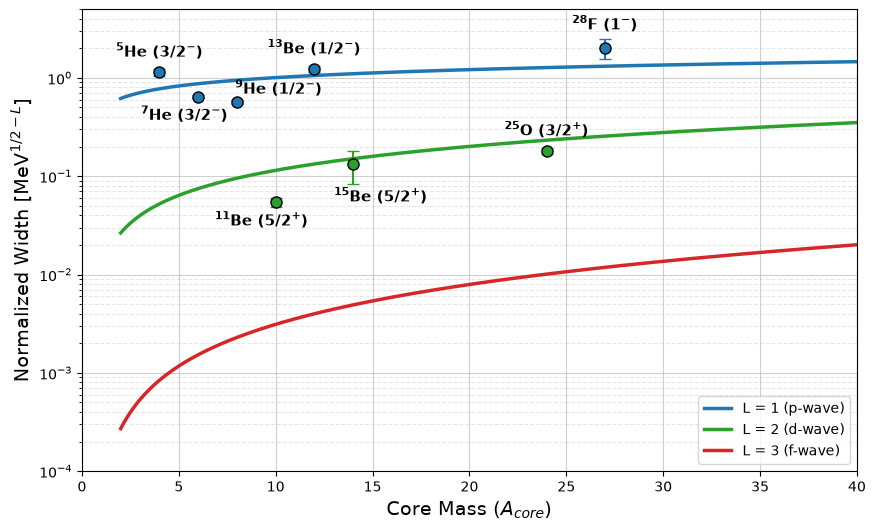}
    \caption{Normalized resonance width $\Gamma_L / E_r^{L+1/2}$ as a function of core mass $A_{\text{core}}$ for near-threshold single-neutron resonances. Solid traces represent theoretical square-well baselines for $p$-wave ($L=1$), $d$-wave ($L=2$), and $f$-wave ($L=3$) states using Eq.~(\ref{eq:master_formula}) with $R = R_{\text{eff}} = 1.14(A_{\text{core}}^{1/3} + 1)$ fm and $\mu = \frac{A_{\text{core}}}{A_{\text{core}}+1} m_n$. Experimental data points include uncertainties for $^{5}$He, $^{7}$He, $^{9}$He, $^{11}$Be, $^{13}$Be, $^{15}$Be, $^{25}$O, and $^{28}$F.}
    \label{fig:normalized_width}
\end{figure*}

The properties of Eq.~(\ref{eq:master_formula}) allow for the construction of a robust phenomenological baseline. By factoring out the kinematic energy dependence, we obtain the normalized width $\Gamma_L / E_r^{L+1/2}$, which is governed entirely by the geometric bounds and the reduced mass of the system.  This normalized ratio is also consistent with Wigner's prediction of the cross section behavior near threshold \cite{Wigner:1948zz}. As shown in Fig.~\ref{fig:normalized_width}, this normalized trace serves as an analytical reference across the nuclear chart. Because the variables in this ratio are deterministic functions of the core mass, the threshold decay behavior of a single-neutron state is bounded by the universal properties of the potential. Consequently, experimental deviations from this trace provide immediate, quantitative indicators of internal structure phenomena or spectroscopic fragmentation.

In conclusion, we have derived a scale-invariant, square-well benchmark for near-threshold single-neutron resonances in $L>0$ partial waves. For a specified channel radius, the resulting width prediction relies solely on geometry, angular momentum, and resonance energy, effectively erasing dependence on internal radial excitation in the sharp-boundary limit. Application to an expanded dataset of observed $p$- and $d$-wave resonances demonstrates that these baseline predictions successfully capture the physical scale of the experimental widths. Furthermore, we demonstrated that realistic Woods-Saxon diffuseness explicitly breaks this discrete scale invariance. This provides a geometric mechanism for width suppression, highlighting how surface effects and internal probability dilution introduce predictable scatter around the idealized universal limit.

Looking ahead, this geometric framework can serve as a vital diagnostic tool for future investigations of neutron-rich nuclei near the dripline. As high-intensity radioactive ion beam facilities extend our experimental reach into heavier, more exotic mass regions—such as the neutron-rich isotopes of calcium and tin—the single-particle strength may fragment due to dense level structures and emergent many-body complexity \cite{Michel:2008pt}. In this uncharted regime, isolating universal continuum kinematics from true structural phenomena—such as extended neutron halos, core excitations, pairing phase transitions, or enhanced surface diffuseness—is important. While recent phenomenological studies have modeled these emission widths using empirical boundary parameters \cite{Delion:2023vse}, our universal predictions provide a simple, parameter-free geometric baseline for this purpose. 

Because the near-threshold width is overwhelmingly dominated by the $E_r^{L+1/2}$ kinematic scaling, the geometric baselines for different partial waves are separated by orders of magnitude. Consequently, this framework can act as a spectroscopic tool. In cases where experimental resolution or statistics are insufficient to extract angular distributions, plotting the measured energy and width against these universal traces can establish or strongly constrain the orbital angular momentum $L$ of the state. As future experiments resolve new near-threshold resonances, systematic deviations from this geometric trace will provide a clear roadmap for charting the evolution of nuclear structure at the limits of stability.

\begin{acknowledgments}
{\it Acknowledgments:} We are grateful for helpful discussions with Witek Nazarewicz and many members of the Nuclear Lattice Effective Field Theory Collaboration. We acknowledge the following funding and computing resources: M.K. and Y.K. acknowledge the support from the Institute for Basic Science (IBS) of the Republic of Korea (Grants No. IBS-R031-D1).  This work was supported in part by the National Research Foundation of Korea funded by Ministry of Science and ICT (RS-2024-00436392); Computational resources were provided by the National Supercomputing Center of Korea with supercomputing resources 
including technical support (KSC-2024-CRE-0335, KSC-2025-CHA-0004).
D.L. and Y.-Z.M. acknowledge U.S. Department of Energy grants DE-SC0013365, DE-SC0023175, DE-SC0026198, DE-SC0023658; U.S. National Science Foundation grant PHY-2310620; as well as the Oak Ridge Leadership Computing Facility computing resources through the INCITE award ``Ab-initio nuclear structure and nuclear reactions'' and Michigan State University's Institute for Cyber-Enabled Research and High-Performance Computing Center.
The work of Y.-H. Song was supported by the Rare Isotope Science Project of the Institute for Basic Science (IBS-I001-01);  National Research Foundation of Korea (NRF) funded by the Ministry of Science and ICT (2013M7A1A1075764,RS-2024-00436392); Computational resources were provided by the National Supercomputing Center of Korea with supercomputing resources 
including technical support (KSC-2024-CRE-0256).
H.-W.H. was supported in part by the Deutsche Forschungsgemeinschaft (DFG, German Research Foundation) - Project ID 279384907 - SFB 1245 and by the BMFTR Contract No. 05P24RDB.
\end{acknowledgments}

\bibliography{References}

@article{Higa:2008dn,
    author = "Higa, R. and Hammer, H. -W. and van Kolck, U.",
    title = "{alpha alpha Scattering in Halo Effective Field Theory}",
    eprint = "0802.3426",
    archivePrefix = "arXiv",
    primaryClass = "nucl-th",
    doi = "10.1016/j.nuclphysa.2008.06.003",
    journal = "Nucl. Phys. A",
    volume = "809",
    pages = "171--188",
    year = "2008"
}

@article{Schmickler:2019ewl,
    author = "Schmickler, C. H. and Hammer, H. -W. and Volosniev, A. G.",
    title = "{Universal physics of bound states of a few charged particles}",
    eprint = "1904.00913",
    archivePrefix = "arXiv",
    primaryClass = "nucl-th",
    doi = "10.1016/j.physletb.2019.135016",
    journal = "Phys. Lett. B",
    volume = "798",
    pages = "135016",
    year = "2019"
}

@article{Mochizuki:2024dbf,
    author = "Mochizuki, Shunta and Nishida, Yusuke",
    title = "{Universal bound states and resonances with Coulomb plus short-range potentials}",
    eprint = "2408.06011",
    archivePrefix = "arXiv",
    primaryClass = "nucl-th",
    doi = "10.1103/PhysRevC.110.064001",
    journal = "Phys. Rev. C",
    volume = "110",
    number = "6",
    pages = "064001",
    year = "2024"
}

@article{Bringas:2004zz,
    author = "Bringas, F. and Yamashita, M. T. and Frederico, T.",
    title = "{Triatomic continuum resonances for large negative scattering lengths}",
    eprint = "cond-mat/0312291",
    archivePrefix = "arXiv",
    doi = "10.1103/PhysRevA.69.040702",
    journal = "Phys. Rev. A",
    volume = "69",
    pages = "040702",
    year = "2004"
}

@article{Deltuva:2020sdd,
    author = "Deltuva, A.",
    title = "{Energies and widths of Efimov states in the three-boson continuum}",
    eprint = "2101.12654",
    archivePrefix = "arXiv",
    primaryClass = "cond-mat.quant-gas",
    doi = "10.1103/PhysRevC.102.034003",
    journal = "Phys. Rev. C",
    volume = "102",
    number = "3",
    pages = "034003",
    year = "2020"
}

@article{Dietz:2021haj,
    author = {Dietz, Sebastian and Hammer, Hans-Werner and K{\"o}nig, Sebastian and Schwenk, Achim},
    title = "{Three-body resonances in pionless effective field theory}",
    eprint = "2109.11356",
    archivePrefix = "arXiv",
    primaryClass = "nucl-th",
    doi = "10.1103/PhysRevC.105.064002",
    journal = "Phys. Rev. C",
    volume = "105",
    number = "6",
    pages = "064002",
    year = "2022"
}

@article{Hammer:2010fw,
    author = "Hammer, H.-W. and Lee, D.",
    title = "{Causality and the effective range expansion}",
    doi = "10.1016/j.aop.2010.06.006",
    journal = "Annals Phys.",
    volume = "325",
    pages = "2212--2233",
    year = "2010",
    eprint = "1002.4603",
    archivePrefix = "arXiv",
    primaryClass = "nucl-th"
}

@article{Tilley:2002vg,
    author = "Tilley, D. R. and Cheves, C. M. and Godwin, J. L. and Hale, G. M. and Hayes, H. M. and Kelley, J. H. and Sheu, C. G. and Weller, H. R.",
    title = "{Energy levels of light nuclei $A=5$, 6, 7}",
    doi = "10.1016/S0375-9474(02)00597-3",
    journal = "Nucl. Phys. A",
    volume = "708",
    pages = "3--163",
    year = "2002"
}

@article{Kondo:2010,
    author = "Kondo, Y. and others",
    title = "{Low-lying resonant states in $^{13}$Be}",
    doi = "10.1016/j.physletb.2010.05.031",
    journal = "Phys. Lett. B",
    volume = "690",
    pages = "245--249",
    year = "2010"
}

@article{Snyder:2013,
    author = "Snyder, J. and others",
    title = "{First observation of the unbound nucleus $^{15}$Be}",
    doi = "10.1103/PhysRevC.88.031303",
    journal = "Phys. Rev. C",
    volume = "88",
    pages = "031303",
    year = "2013"
}

@article{Kondo:2016vzi,
    author = "Kondo, Y. and others",
    title = "{Nucleus $^{26}$O: A Barely Unbound System beyond the Drip Line}",
    doi = "10.1103/PhysRevLett.116.102503",
    journal = "Phys. Rev. Lett.",
    volume = "116",
    pages = "102503",
    year = "2016"
}

@article{Hammer:2009zh,
    author = "Hammer, H. -W. and Lee, Dean",
    title = "{Causality and universality in low-energy quantum scattering}",
    eprint = "0907.1763",
    archivePrefix = "arXiv",
    primaryClass = "nucl-th",
    reportNumber = "HISKP-TH-09-25",
    doi = "10.1016/j.physletb.2009.10.033",
    journal = "Phys. Lett. B",
    volume = "681",
    pages = "500--503",
    year = "2009"
}

@article{Phillips:1996ae,
    author = "Phillips, Daniel R. and Cohen, Thomas D.",
    title = "{How short is too short? Constraining contact interactions in nucleon-nucleon scattering}",
    eprint = "nucl-th/9607048",
    archivePrefix = "arXiv",
    reportNumber = "DOE-ER-40762-092, UMD-PP-97-008",
    doi = "10.1016/S0370-2693(96)01411-6",
    journal = "Phys. Lett. B",
    volume = "390",
    pages = "7--12",
    year = "1997"
}

@article{Wigner:1955zz,
    author = "Wigner, Eugene P.",
    title = "{Lower Limit for the Energy Derivative of the Scattering Phase Shift}",
    doi = "10.1103/PhysRev.98.145",
    journal = "Phys. Rev.",
    volume = "98",
    pages = "145--147",
    year = "1955"
}

@article{Sun:2026ntq,
  author = "{Sun, Y. L. and others}",
  title = "{Observation of a broad $p$-wave resonant state in $^{9}$He}",
  journal = {arXiv preprint},
  eprint = "2602.07810",
  archivePrefix = "arXiv",
  primaryClass = "nucl-ex",
  year = "2026"
}

@article{Kelley:2012zz,
    author = "Kelley, J. H. and Kwan, E. and Purcell, J. E. and Sheu, C. G. and Weller, H. R.",
    title = "{Energy levels of light nuclei $A=11$}",
    doi = "10.1016/j.nuclphysa.2012.01.010",
    journal = "Nucl. Phys. A",
    volume = "880",
    pages = "88--195",
    year = "2012"
}

@article{Hoffman:2008szc,
    author = "Hoffman, C. R. and others",
    title = "{Determination of the $N=16$ halo shell closure at the oxygen drip line}",
    eprint = "0801.3855",
    archivePrefix = "arXiv",
    primaryClass = "nucl-ex",
    doi = "10.1103/PhysRevLett.100.152502",
    journal = "Phys. Rev. Lett.",
    volume = "100",
    pages = "152502",
    year = "2008"
}

@article{Corsi:2019bto,
    author = "Corsi, A. and others",
    title = "{Structure of $^{13}$Be studied in proton knockout from $^{14}$B}",
    eprint = "1908.09459",
    archivePrefix = "arXiv",
    primaryClass = "nucl-ex",
    doi = "10.1016/j.physletb.2019.134843",
    journal = "Phys. Lett. B",
    volume = "797",
    pages = "134843",
    year = "2019"
}

@article{Jensen:2004zz,
    author = "Jensen, A. S. and Riisager, K. and Fedorov, D. V. and Garrido, E.",
    title = "{Structure and reactions of quantum halos}",
    doi = "10.1103/RevModPhys.76.215",
    journal = "Rev. Mod. Phys.",
    volume = "76",
    pages = "215--261",
    year = "2004"
}

@article{Michel:2008pt,
    author = "Michel, N. and Nazarewicz, W. and Ploszajczak, M. and Vertse, T.",
    title = "{Shell Model in the Complex Energy Plane}",
    eprint = "0810.2728",
    archivePrefix = "arXiv",
    primaryClass = "nucl-th",
    doi = "10.1088/0954-3899/36/1/013101",
    journal = "J. Phys. G",
    volume = "36",
    pages = "013101",
    year = "2009"
}

@article{Braaten:2004rn,
    author = "Braaten, Eric and Hammer, H. -W.",
    title = "{Universality in few-body systems with large scattering length}",
    eprint = "cond-mat/0410417",
    archivePrefix = "arXiv",
    reportNumber = "INT-PUB-04-27",
    doi = "10.1016/j.physrep.2006.03.001",
    journal = "Phys. Rept.",
    volume = "428",
    pages = "259--390",
    year = "2006"
}

@article{Lane:1958vxu,
    author = "Lane, A. M. and Thomas, R. G.",
    title = "{R-Matrix Theory of Nuclear Reactions}",
    doi = "10.1103/RevModPhys.30.257",
    journal = "Rev. Mod. Phys.",
    volume = "30",
    pages = "257--353",
    year = "1958"
}

@article{nussenzveig1959poles,
  title={The poles of the S-matrix of a rectangular potential well of barrier},
  author={Nussenzveig, Herch Moys{\'e}s},
  journal={Nuclear Physics},
  volume={11},
  pages={499--521},
  year={1959},
  publisher={Elsevier}
}

@article{Gurvitz:1986uv,
    author = "Gurvitz, S. A. and Kaelbermann, German",
    title = "{The Decay Width and the Shift of a Quasistationary State}",
    reportNumber = "WIS-86/51-Ph",
    doi = "10.1103/PhysRevLett.59.262",
    journal = "Phys. Rev. Lett.",
    volume = "59",
    pages = "262",
    year = "1987"
}

@article{Wigner:1948zz,
    author = "Wigner, Eugene P.",
    title = "{On the Behavior of Cross Sections Near Thresholds}",
    doi = "10.1103/PhysRev.73.1002",
    journal = "Phys. Rev.",
    volume = "73",
    pages = "1002--1009",
    year = "1948"
}

@article{Efimov:1970zz,
    author = "Efimov, V.",
    title = "{Energy levels arising from resonant two-body forces in a three-body system}",
    doi = "10.1016/0370-2693(70)90349-7",
    journal = "Phys. Lett. B",
    volume = "33",
    pages = "563--564",
    year = "1970"
}

@article{Efimov:1971zz,
    author = "Efimov, V. N.",
    title = "{Weakly-Bound State of 3 Resonantly-Interacting Particles}",
    journal = "Sov. J. Nucl. Phys.",
    volume = "12",
    pages = "589",
    year = "1971"
}

@article{efimov2009giant,
  title={Giant trimers true to scale},
  author={Efimov, Vitaly},
  journal={Nature Physics},
  volume={5},
  number={8},
  pages={533--534},
  year={2009},
  publisher={Nature Publishing Group UK London}
}

@article{naidon2017efimov,
  title={Efimov physics: a review},
  author={Naidon, Pascal and Endo, Shimpei},
  journal={Reports on Progress in Physics},
  volume={80},
  number={5},
  pages={056001},
  year={2017},
  publisher={IOP Publishing}
}

@article{hammer2010efimov,
  title={Efimov states in nuclear and particle physics},
  author={Hammer, Hans-Werner and Platter, Lucas},
  journal={Annual Review of Nuclear and Particle Science},
  volume={60},
  number={1},
  pages={207--236},
  year={2010},
  publisher={Annual Reviews}
}

@article{kievsky2021efimov,
  title={Efimov physics and connections to nuclear physics},
  author={Kievsky, A and Gattobigio, M and Girlanda, L and Viviani, M},
  journal={Annual Review of Nuclear and Particle Science},
  volume={71},
  number={1},
  pages={465--490},
  year={2021},
  publisher={Annual Reviews}
}

@article{kraemer2006evidence,
  title={Evidence for Efimov quantum states in an ultracold gas of caesium atoms},
  author={Kraemer, Tobias and Mark, Manfred and Waldburger, Philipp and Danzl, Johann G and Chin, Cheng and Engeser, Bastian and Lange, Almar D and Pilch, Karl and Jaakkola, Antti and N{\"a}gerl, H-C and others},
  journal={Nature},
  volume={440},
  number={7082},
  pages={315--318},
  year={2006},
  publisher={Nature Publishing Group UK London}
}

@article{kunitski2015observation,
  title={Observation of the Efimov state of the helium trimer},
  author={Kunitski, Maksim and Zeller, Stefan and Voigtsberger, J{\"o}rg and Kalinin, Anton and Schmidt, Lothar Ph H and Sch{\"o}ffler, Markus and Czasch, Achim and Sch{\"o}llkopf, Wieland and Grisenti, Robert E and Jahnke, Till and others},
  journal={Science},
  volume={348},
  number={6234},
  pages={551--555},
  year={2015},
  publisher={American Association for the Advancement of Science}
}

@article{wang2012origin,
  title={Origin of the three-body parameter universality in Efimov physics},
  author={Wang, Jia and D’Incao, Jose P and Esry, BD and Greene, Chris H},
  journal={Physical review letters},
  volume={108},
  number={26},
  pages={263001},
  year={2012},
  publisher={APS}
}

@article{braaten2007efimov,
  title={Efimov physics in cold atoms},
  author={Braaten, Eric and Hammer, H-W},
  journal={Annals of Physics},
  volume={322},
  number={1},
  pages={120--163},
  year={2007},
  publisher={Elsevier}
}

@article{nishida2012weakly,
  title={Weakly bound molecules trapped with discrete scaling symmetries},
  author={Nishida, Yusuke and Lee, Dean},
  journal={Physical Review A—Atomic, Molecular, and Optical Physics},
  volume={86},
  number={3},
  pages={032706},
  year={2012},
  publisher={APS}
}

@article{nishida2011liberating,
  title={Liberating Efimov physics from three dimensions},
  author={Nishida, Yusuke and Tan, Shina},
  journal={Few-Body Systems},
  volume={51},
  number={2},
  pages={191--206},
  year={2011},
  publisher={Springer}
}

@article{Delion:2023vse,
    author = "Delion, D. S. and Ghinescu, S.",
    title = "{Systematics of neutron emission}",
    eprint = "2311.08180",
    archivePrefix = "arXiv",
    primaryClass = "nucl-th",
    doi = "10.1088/1361-6471/add02e",
    journal = "J. Phys. G",
    volume = "52",
    number = "5",
    pages = "055105",
    year = "2025"
}

@article{PhysRevLett.124.152502,
  title = {Extending the Southern Shore of the Island of Inversion to $^{28}\mathrm{F}$},
  author = {Revel, A. and others},
  collaboration = {SAMURAI21 collaboration},
  journal = {Phys. Rev. Lett.},
  volume = {124},
  issue = {15},
  pages = {152502},
  numpages = {7},
  year = {2020},
  month = {Apr},
  publisher = {American Physical Society},
  doi = {10.1103/PhysRevLett.124.152502},
}

@article{CAO201246,
title = "{Recoil proton tagged knockout reaction for $^{8}$He}",
journal = "Phys. Lett. B",
volume = {707},
number = {1},
pages = {46-51},
year = {2012},
issn = {0370-2693},
doi = {https://doi.org/10.1016/j.physletb.2011.12.009},
author = {Z.X. Cao and others},
}

\clearpage
\onecolumngrid
\begin{center}
\textbf{\large Supplemental Material}
\end{center}
\vspace{4mm}

\setcounter{equation}{0}
\setcounter{figure}{0}
\setcounter{table}{0}
\setcounter{section}{0}
\renewcommand{\theequation}{S\arabic{equation}}
\renewcommand{\thefigure}{S\arabic{figure}}
\renewcommand{\thesection}{S\arabic{section}}
\setcounter{secnumdepth}{3}

\section{Discrete Scale Invariance and Node Independence}

The independence of the threshold resonance width on the internal radial excitation is a consequence of a discrete scale symmetry unique to the spherical square well. This symmetry allows the properties of one radial excitation to be mapped directly onto another without altering the underlying dynamics. To demonstrate this, consider the zero-energy radial Schr\"{o}dinger equation strictly within a nuclear boundary, $r \le R$. We parameterize a generic finite-range potential by a peak depth $-V_0$ and a dimensionless shape function $f(r/R)$:
\begin{equation}
\frac{d^2u}{dr^2} - \frac{L(L+1)}{r^2}u + K_0^2 f\left(\frac{r}{R}\right) u = 0
\end{equation}
where $K_0 = \sqrt{2\mu V_0}$ is the maximum internal momentum. At the zero-energy threshold, the exterior wave function must match the purely kinematic asymptotic form $u_{ext}(r) \propto r^{-L}$. Therefore, boundary matching requires that the dimensionless logarithmic derivative evaluated at the surface $r=R$ is $-L$:
\begin{equation}
\left. \frac{r}{u} \frac{du}{dr} \right|_{r=R} = -L
\end{equation}
To test for discrete scale invariance, we transform the system into the dimensionless phase coordinate $\rho = K_0 r$. The boundary is correspondingly shifted to $\rho_R = K_0 R$. Because the dimensionless logarithmic derivative operator retains its mathematical form under coordinate substitution, the boundary condition translates directly into the scaled coordinates:
\begin{equation}
\left. \frac{\rho}{u} \frac{du}{d\rho} \right|_{\rho=\rho_R} = -L
\end{equation}
Substituting $r = \rho/K_0$ into the Schr\"{o}dinger equation, the internal differential equation becomes:
\begin{equation}
\frac{d^2u}{d\rho^2} - \frac{L(L+1)}{\rho^2}u + f\left(\frac{\rho}{\rho_R}\right) u = 0
\end{equation}

\subsection{The Square Well: Emergence of Discrete Symmetry}

For a uniform interior---the spherical square well---the shape function is a constant $f(x) = 1$. The boundary parameter $\rho_R$ vanishes from the differential equation, yielding an invariant form:
\begin{equation}
\frac{d^2u}{d\rho^2} - \frac{L(L+1)}{\rho^2}u + u = 0
\end{equation}
The solution to this equation, $u(\rho) \propto \rho j_L(\rho)$, is decoupled from the macroscopic boundary $R$ and the specific radial excitation node number $n$. The boundary condition $\left. \frac{\rho}{u} \frac{du}{d\rho} \right|_{\rho=\rho_R} = -L$ is satisfied only at specific, discrete nodes. To demonstrate this, we use the spherical Bessel recurrence relation, $j_L'(z) = j_{L-1}(z) - \frac{L+1}{z}j_L(z)$, to evaluate the internal logarithmic derivative:
\begin{equation}
\frac{\rho}{u}\frac{du}{d\rho} = \rho \frac{j_{L-1}(\rho)}{j_L(\rho)} - L
\end{equation}
Equating this to the $-L$ threshold boundary condition at $\rho = \rho_R$ leaves only the condition $j_{L-1}(\rho_R) = 0$. This requires the dimensionless boundary to sit precisely at the nodes of the Bessel function, $\rho_R^{(n)} = K_n R = \chi_{n,L-1}$.

To explicitly define the discrete scale invariance, consider mapping an internal state with $n$ nodes to a highly excited state with $n'$ nodes within the same fixed physical radius $R$. Because the differential equation is completely blind to $\rho_R$, the internal momentum $K_n$ can be dilated by a specific, discrete scale factor $\lambda$:
\begin{equation}
K_n \to K_{n'} = \lambda K_n \quad \text{where} \quad \lambda = \frac{\chi_{n', L-1}}{\chi_{n, L-1}}
\end{equation}
Under this discrete scale transformation, the dimensionless phase coordinate stretches ($\rho \to \lambda \rho$), and the boundary matching point maps perfectly from one root to another:
\begin{equation}
\rho_R^{(n)} \to \lambda \rho_R^{(n)} = \left( \frac{\chi_{n', L-1}}{\chi_{n, L-1}} \right) \chi_{n, L-1} = \chi_{n', L-1} = \rho_R^{(n')}.
\end{equation}
The functional form of the wave function $u(\rho) \propto \rho j_L(\rho)$ is structurally untouched by this scaling. This transformation compresses the internal wavelength by the amount needed to fit $n'$ nodes into the original radius $R$ without altering the surface amplitude or the slope. Because both the initial and transformed boundaries are exact nodes of $j_{L-1}$, the logarithmic derivative at the fixed physical surface $R$ remains identically $-L$. It is this exact discrete mapping that physically decouples the internal radial excitation from the external decay dynamics and the observable threshold width.

While discrete scale invariance is a mathematical property at exactly zero energy, its physical consequences extend to finite, near-threshold energies ($E_r > 0$). In this regime ($k_r R \ll 1$), the resonance width is governed by the effective range expansion:
\begin{equation}
\Gamma_L = \frac{2k_r^{2L+1}}{\mu |r_L|}\left[1+O(k_r^2)\right].
\end{equation}
In Sec.~\ref{sec:effective_range}, we show that $r_L$ is completely independent of the radial excitation number for the square-well potential.

\subsection{Symmetry Breaking in General Potentials}

This discrete mapping fails for any arbitrary potential where the internal depth varies with distance ($f$ is not constant). When fitting a higher radial excitation (larger $n$) into the fixed radius $R$, the internal momentum $K_0$ must increase to satisfy the boundary matching, which increases $\rho_R = K_0 R$. For a generic potential, $\rho_R$ remains inside the argument of the shape function $f(\rho/\rho_R)$. Consequently, the structure of the differential equation:
\begin{equation}
\frac{d^2u}{d\rho^2} - \frac{L(L+1)}{\rho^2}u + f\left(\frac{\rho}{\rho_R}\right) u = 0
\end{equation}
changes for every node $n$. A coordinate rescaling can no longer map the $n$ state to the $n'$ state because the shape of the potential as a function of $\rho$ is altered. The ratio of the internal normalization integral to the exterior boundary amplitude becomes dependent on $n$, breaking the discrete scale invariance and restoring the width's dependence on the internal structure.

\section{Effective Range and the square-well Integral}\label{sec:effective_range}

The generalized Bethe integral formula for a finite-range interaction (range $R$) connects the effective range parameter $r_L$ to the zero-energy wave function $u_L^{(0)}(r)$ \cite{Hammer:2010fw}:
\begin{equation}
    r_L = b_L(R) - 2 \int_{0}^{R} \left[ u_L^{(0)}(r) \right]^2 dr.
\end{equation}
Here, $b_L(R)$ is determined by the exterior solution. For a zero-energy resonance where the scattering length diverges ($|a_L| \to \infty$), the causality bound function evaluates to:
\begin{equation}
    b_L(R) = -\frac{2\Gamma(L-1/2)\Gamma(L+1/2)}{\pi} \left(\frac{R}{2}\right)^{1-2L}.
\end{equation}
For a spherical square well $V(r) = -V_0 \theta(R-r)$, the zero-energy internal wave function is $u_{in}(r) = A r j_L(\kappa r)$, where $\kappa = \sqrt{2\mu V_0}$. The exterior zero-energy wave function must match the correct asymptotic normalization, 
\begin{equation}
    u_{ext}(r) = \frac{\Gamma(L+1/2)}{\sqrt{\pi}} \left(\frac{r}{2}\right)^{-L}.
\end{equation}
We define the boundary amplitude as $U_R \equiv u_{ext}(R)$. 

Matching the logarithmic derivatives at $r=R$ yields the resonance condition:
\begin{equation}
    j_{L-1}(\kappa R) = 0.
\end{equation}
Matching the wave function amplitudes directly yields $A R j_L(\kappa R) = U_R$. To evaluate the internal normalization $I = 2 \int_0^R [u_{in}(r)]^2 dr$, we substitute $A$ and use the spherical Bessel integral identity:
\begin{equation}
    \int_0^R r^2 j_L^2(\kappa r) dr = \frac{R^3}{2} \left[ j_L^2(\kappa R) - j_{L-1}(\kappa R)j_{L+1}(\kappa R) \right].
\end{equation}
Applying the resonance condition $j_{L-1}(\kappa R) = 0$, the integration evaluates analytically. Substituting the bound amplitude $U_R$ provides a simple geometric result:
\begin{equation}
    I = 2 \left( \frac{U_R}{R j_L(\kappa R)} \right)^2 \left( \frac{R^3}{2} j_L^2(\kappa R) \right) = U_R^2 R.
\end{equation}
Evaluating the ratio of this integral correction to the bound term $b_L(R)$ yields a geometric relationship entirely independent of the potential depth $V_0$:
\begin{equation}
    \frac{I}{|b_L(R)|} = L - \frac{1}{2}.
\end{equation}
Consequently, the effective range for the square well is $r_L = -(L + 1/2)|b_L(R)|$. We note that this formula relies entirely on boundary properties and is completely independent of the radial excitation number.

\section{Geometric Origin of the Node Independence}
\label{sec:geometrical}

Section~\ref{sec:effective_range} showed algebraically, through the
identity
\begin{equation}
    I \equiv 2\int_0^R u^2(r)\,dr = R U_R^2 ,
\end{equation}
that the square-well contribution to the near-threshold effective range is
independent of the radial excitation number. Here we give a complementary
geometric interpretation. The key point is that, at zero energy, the
normalization integral inside a square well potential can be converted exactly
into a boundary term. This boundary form makes the cancellation of the
internal momentum scale manifest.

We begin with the zero-energy radial equation inside the well,
\begin{equation}
    u''(r)-\frac{L(L+1)}{r^2}u(r)+K_0^2 u(r)=0,
    \qquad 0<r<R ,
\end{equation}
where \(K_0=\sqrt{2\mu V_0}\). First consider the scale-invariant part of
this equation, obtained by setting \(K_0=0\). The corresponding
one-dimensional Lagrangian density may be written as
\begin{equation}
    \mathcal{L}_{0}
    =
    \frac{1}{2}\bigl(u'\bigr)^2
    +
    \frac{L(L+1)}{2r^2}u^2 .
\end{equation}
This action is invariant under the scale transformation
\begin{equation}
    r\to \lambda r,
    \qquad
    u(r)\to \lambda^{1/2}u(\lambda r),
\end{equation}
or, equivalently, under the infinitesimal variation
\begin{equation}
    \delta r = \epsilon r,
    \qquad
    \delta u = \epsilon\left(r u' + \frac12 u\right).
\end{equation}
The associated dilation current is
\begin{equation}
    D_0(r)
    =
    r\bigl(u'\bigr)^2
    -
    u u'
    -
    \frac{L(L+1)}{r}u^2 .
\end{equation}
For solutions of the \(K_0=0\) equation, this current is conserved,
\(dD_0/dr=0\).

The square well adds the uniform deformation
\begin{equation}
    \mathcal{L}
    =
    \mathcal{L}_0
    -
    \frac{1}{2}K_0^2 u^2 .
\end{equation}
This term introduces the scale \(K_0\) and breaks the continuous scale
symmetry. The corresponding modified dilation current inside the well is
\begin{equation}
    D(r)
    =
    r\bigl(u'\bigr)^2
    -
    u u'
    -
    \frac{L(L+1)}{r}u^2
    +
    K_0^2 r u^2 .
\end{equation}
Using the full square-well equation of motion, one finds the exact
identity
\begin{equation}
    \frac{dD}{dr}
    =
    2K_0^2 u^2 .
    \label{eq:D_identity}
\end{equation}
This is the scale Ward identity for the uniformly deformed problem.

Integrating Eq.~\eqref{eq:D_identity} from the origin to the well radius
gives
\begin{equation}
    2K_0^2\int_0^R u^2(r)\,dr
    =
    D(R^-)-D(0),
\end{equation}
where \(D(R^-)\) denotes the current evaluated just inside the square
well. For the regular solution at the origin, \(u(r)\sim r^{L+1}\), all
terms in \(D(r)\) vanish as \(r\to 0\). Hence \(D(0)=0\), and the
interior normalization is determined entirely by the interior boundary
current,
\begin{equation}
    \int_0^R u^2(r)\,dr
    =
    \frac{D(R^-)}{2K_0^2}.
    \label{eq:norm_boundary}
\end{equation}

At threshold, the exterior zero-energy wave function is proportional to
\(r^{-L}\). Therefore the logarithmic derivative at the boundary is fixed
kinematically,
\begin{equation}
    u(R)=U_R,
    \qquad
    u'(R)=-\frac{L}{R}U_R .
\end{equation}
Before evaluating the current explicitly, we note why the scale-invariant
part must vanish at the boundary. For the \(K_0=0\) equation, the two
independent zero-energy solutions are
\begin{equation}
    u(r)\propto r^{L+1},
    \qquad
    u(r)\propto r^{-L}.
\end{equation}
Both are pure scale eigenfunctions. For a general power-law solution
\(u(r)\propto r^\alpha\), one has \(ru'/u=\alpha\), and therefore
\begin{equation}
    D_0(r)
    =
    \frac{u^2}{r}
    \left[
        \alpha^2-\alpha-L(L+1)
    \right].
\end{equation}
The allowed zero-energy exponents obey
\begin{equation}
    \alpha(\alpha-1)=L(L+1),
\end{equation}
so \(D_0(r)=0\) on either pure scaling branch. A nonzero dilation current
can arise only from a mixture of the two branches.

Equivalently, since the square well has a finite jump but no
delta-function shell at \(r=R\), both \(u\) and \(u'\) are continuous
across the boundary. The current \(D_0\), which depends only on \(u\) and
\(u'\), is therefore continuous at \(R\). Outside the well, the threshold
solution is the pure scale eigenfunction \(u_{\rm ext}\propto r^{-L}\),
for which \(D_0=0\) identically. Hence
\begin{equation}
    D_0(R^-)=D_0(R^+)=0 .
\end{equation}
This shows that the vanishing of \(D_0(R)\) is not an accidental
algebraic cancellation; it follows from matching to a pure exterior
scaling branch.

Substituting the boundary condition directly gives the same result,
\begin{align}
    D_0(R^-)
    &=
    R\left(-\frac{L}{R}U_R\right)^2
    -
    U_R\left(-\frac{L}{R}U_R\right)
    -
    \frac{L(L+1)}{R}U_R^2
    \nonumber \\
    &=0 .
\end{align}
Thus the interior boundary current receives only the contribution from
the uniform square-well deformation,
\begin{equation}
    D(R^-)
    =
    K_0^2 R U_R^2 .
\end{equation}
Equation~\eqref{eq:norm_boundary} then yields
\begin{equation}
    \int_0^R u^2(r)\,dr
    =
    \frac{1}{2K_0^2}
    \left(K_0^2 R U_R^2\right)
    =
    \frac{R}{2}U_R^2 .
\end{equation}
Therefore the integral correction appearing in
Sec.~\ref{sec:effective_range},
\begin{equation}
    I \equiv 2\int_0^R u^2(r)\,dr,
\end{equation}
is
\begin{equation}
    I = R U_R^2 ,
\end{equation}
which is precisely the result obtained there from the explicit spherical
Bessel integral.

The dependence on the internal momentum scale \(K_0\) cancels between
the scale-breaking Ward identity and the interior boundary current.
Since different radial excitations of the square well correspond to
different discrete values of \(K_0R=\chi_{n,L-1}\), this cancellation
explains why the threshold effective range, and therefore the
near-threshold width coefficient, is independent of the radial node
number.
\section{R-Matrix Width and Penetrability}

Near threshold, the scattering phase shift is expanded as $k^{2L+1}\cot\delta_L\approx\frac{1}{2}r_L(k^2-k_r^2)$. Evaluating this at the complex S-matrix pole $E_{pole}\approx E_r-i\Gamma_L/2$ yields the resonance width $\Gamma_L=2k_r^{2L+1}/(\mu|r_L|)$.

This result maps directly onto the standard R-matrix formalism \cite{Lane:1958vxu}, which factorizes the width into a structural reduced width and a kinematic penetrability: $\Gamma_L=2P_L(k_r R)\gamma^2$. The penetrability $P_L(kR)$ represents the probability of the neutron tunneling through the centrifugal barrier and is completely determined by the exterior wave functions evaluated at the channel radius $R$. For neutral particles lacking a Coulomb barrier, the regular and irregular solutions are the spherical Bessel and Neumann functions, $F_L(kR)=kR j_L(kR)$ and $G_L(kR)=-kR n_L(kR)$. The penetrability is defined as:
\begin{equation}
    P_L(kR) = \frac{kR}{F_L^2(kR) + G_L^2(kR)} = \frac{1}{kR[j_L^2(kR) + n_L^2(kR)]}.
\end{equation}

In the near-threshold limit ($kR \ll 1$), the irregular Neumann function dominates the denominator. Expanding this expression yields the threshold asymptotic form:
\begin{equation}
    P_L(kR) = \frac{(kR)^{2L+1}}{[(2L-1)!!]^2} \left[ 1 - \frac{(kR)^2}{2L-1} + \mathcal{O}\left((kR)^4\right) \right].
\end{equation}

To extract the reduced width for the square well, we equate the standard R-matrix width utilizing this asymptotic penetrability with the closed-form universal width derived in Eq.~(4) of the main text:
\begin{equation}
    2\gamma^2_{\text{SW}}\frac{(k_r R)^{2L+1}}{[(2L-1)!!]^2} = \frac{\pi(2\mu E_r)^{L+1/2}\left(\frac{R}{2}\right)^{2L-1}}{\mu(L+1/2)\Gamma(L-1/2)\Gamma(L+1/2)}.
\end{equation}

Recalling that $k_r = \sqrt{2\mu E_r}$, the energy and mass terms associated with the threshold kinematics precisely cancel on both sides. Utilizing the semi-integer Gamma function identities $\Gamma(L+1/2) = \sqrt{\pi}(2L-1)!!/2^L$ and $\Gamma(L-1/2) = \sqrt{\pi}(2L-3)!!/2^{L-1}$, solving for $\gamma^2_{\text{SW}}$ isolates the purely geometric structural scale:
\begin{equation}
    \gamma^2_{\text{SW}} = \frac{2L-1}{\mu R^2(2L+1)}.
\end{equation}

\section{Woods-Saxon Diffuse Asymptotics and Node Dependence}

Transitioning to a Woods-Saxon potential $V_{\text{WS}}(r)$ with diffuseness $a$ modifies both the internal normalization and the external barrier tunneling. We define the boundary condition at the inner classical turning point $R_{\text{tp}}$, where the effective potential crosses zero: $V_{\text{WS}}(R_{\text{tp}}) + V_B = 0$, with the surface centrifugal barrier $V_B = \frac{\hbar^2 L(L+1)}{2\mu R_{\text{tp}}^2}$.

Because both the square-well and Woods-Saxon threshold wave functions share the same external $r^{-L}$ asymptotic matching at $R_{\text{tp}}$, the shift in the reduced width is determined by the ratio of their internal normalizations evaluated in the WKB approximation: $\gamma^2_{\text{WS}} = \gamma^2_{\text{SW}} ( I_{\text{SW}} / I_{\text{WS}} ) = \gamma^2_{\text{SW}} ( 1 + \Delta I / I_{\text{SW}} )^{-1}$.

The normalization discrepancy $\Delta I = I_{\text{WS}} - I_{\text{SW}}$ arises primarily from the diffuse surface region. Integrating the difference in local WKB momenta over this boundary yields:
\begin{equation}
    \Delta I = \frac{2a R_{\text{tp}}}{\sqrt{L(L+1)}} \left[ 1 - \sqrt{\frac{V_B}{V_0}} \ln\left(\sqrt{\frac{V_0}{V_B}}\right) + \mathcal{O}\left( \sqrt{\frac{V_B}{V_0}} \right) \right].
\end{equation}
For a zero-energy state, the interior matching momentum $K_0 = \sqrt{2\mu V_0}/\hbar$ is constrained by the boundary condition to $K_0 = \chi_{n, L-1} / R_{\text{tp}}$, where $\chi_{n, L-1}$ is the $(n+1)$-th root of $j_{L-1}(z)$. Dividing $\Delta I$ by the square-well normalization $I_{\text{SW}} \approx R_{\text{tp}}/K_0$ yields the normalized internal inflation:
\begin{equation}
    \frac{\Delta I}{I_{\text{SW}}} \approx \frac{2a \chi_{n, L-1}}{R_{\text{tp}}\sqrt{L(L+1)}}.
\end{equation}
Substituting this back into the reduced width formula yields the geometric suppression factor presented in Eq.~(5). Because the spherical Bessel root $\chi_{n, L-1}$ increases monotonically with the radial node number $n$, higher radial excitations require a deeper internal potential ($V_0 \propto \chi_n^2$). This steepens the potential gradient, forcing the wave function to sample the diffuse surface more heavily and severely suppressing $\gamma^2_{\text{WS}}$ relative to nodeless ($n=0$) states.

Concurrently, the diffuse exponential tail spanning $r > R_{\text{tp}}$ modifies the external tunneling probability. We evaluate this enhancement using the WKB barrier penetration integral with the Langer modification ($\lambda = L + 1/2$). The enhancement is governed by the barrier action deficit $\Delta K = \int (\kappa_{\text{SW}} - \kappa_{\text{WS}}) dr$. Substituting $x = r - R_{\text{tp}}$ over the exponential tail, the integral evaluates analytically to:
\begin{equation}
    \Delta K = \frac{\lambda}{2 R_{\text{tp}}^2} \left( R_{\text{tp}} a + a^2 \right) = \frac{\lambda a}{2 R_{\text{tp}}} \left( 1 + \frac{a}{R_{\text{tp}}} \right).
\end{equation}
The physical tunneling probability is modified by $P \propto \exp(-2K)$. The ratio of the true penetrabilities is therefore exponentially enhanced by this action deficit:
\begin{equation}
    \frac{P_{\text{WS}}}{P_{\text{SW}}} = \exp(2\Delta K) \approx \exp\left[ \frac{(L + 1/2) a}{R_{\text{tp}}} \left( 1 + \frac{a}{R_{\text{tp}}} \right) \right].
\end{equation}
Because the internal probability dilation (which suppresses $\gamma^2$) and the external barrier thinning (which enhances $P_L$) depend on geometric boundaries with opposite observable effects, the net structural deviation to the total width $\Gamma_L$ exhibits a strong partial cancellation, ultimately stabilizing the kinematic dominance of the universal square-well baseline.

\section{Analysis of Resonance Width Scaling}

To interpret the experimental resonance widths ($\Gamma$), we utilized a single-particle resonance model where the width is governed by the penetration of the centrifugal barrier. 
The theoretical width is characterized by the energy dependence $\Gamma \propto E_r^{L+1/2}$. This scaling arises from the asymptotic behavior of the free-particle radial wave functions at the resonance energy, where the barrier permeability is dominated by the centrifugal term for a given orbital angular momentum $L$.

\begin{figure*}[h]
    \centering
    \includegraphics[width=13cm]{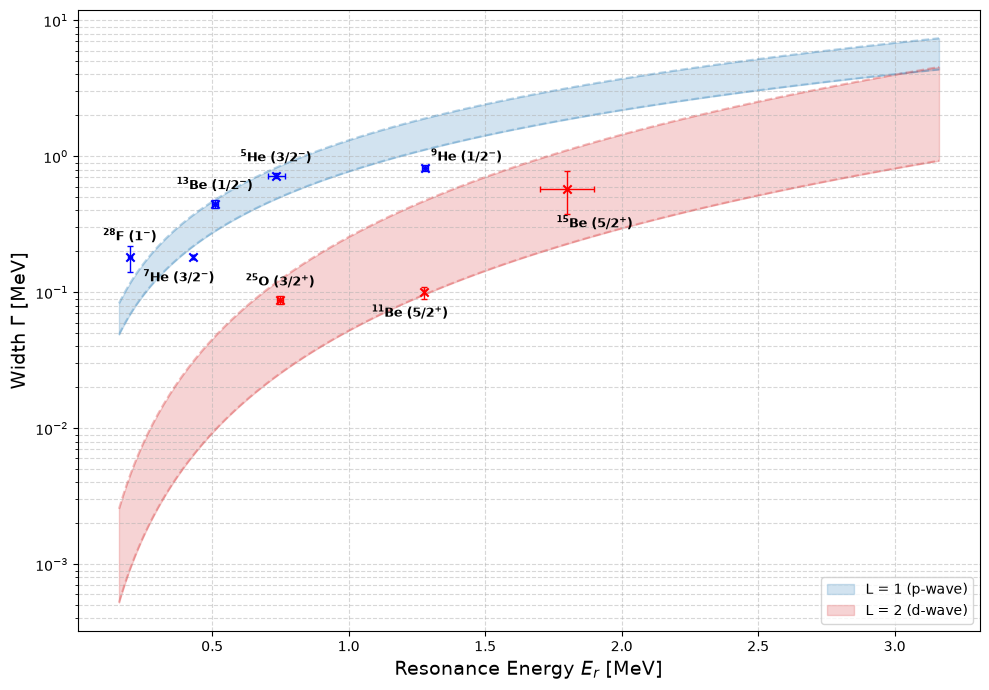}
    \caption{Resonance widths ($\Gamma$) as a function of resonance energy ($E_r$) for the analyzed nuclei. Experimental data are represented by markers with error bars. The blue and red shaded regions represent the theoretical range of widths predicted by the single-particle model for $L=1$ ($p$-wave) and $L=2$ ($d$-wave) states, respectively. The width of these bands reflects the variation in predicted $\Gamma$ across the range of core mass numbers ($A_{\rm core} = 4$ to $27$) present in our dataset, with $A_{\rm core} = 4$ at the bottom of the band and $A_{\rm core} = 27$ at the top of the band.}
    \label{fig:width_Er}
\end{figure*}

Because the theoretical width also depends on the reduced mass ($\mu$) and the effective interaction radius ($R_{\rm eff}$) of the nucleus—both of which evolve with the core mass number ($A_{\rm core}$)—a single-line representation is insufficient for a diverse set of nuclei. 
Consequently, we adopted a shaded band approach to define the ``theoretical range'' for each angular momentum state. 
The bands were generated by calculating the theoretical resonance width for every point in the energy domain using the extreme limits of the core mass numbers found in our dataset ($A_{\rm core} = 4$ to $27$). 
For a fixed $E_r$, the lower boundary represents the predicted width for the lightest core ($A_{\rm core}=4$), while the upper boundary represents the prediction for the heaviest core ($A_{\rm core}=27$). 
This methodology effectively creates an envelope of uncertainty that accounts for the variation in nuclear size and kinematics across our sample, ensuring that any experimental data point falling within these bounds is considered consistent with single-particle behavior.

The shaded band representation serves as a diagnostic tool for identifying collective or non-single-particle nuclear structure effects. 
Data points residing within the $L=1$ or $L=2$ bands indicate that the decay width is primarily determined by the centrifugal barrier penetration, suggesting the resonance is well-described by a valence nucleon orbiting a core. 
Conversely, points lying significantly outside these bands provide evidence for physics beyond the single-particle approximation. 
Data points above the bands may indicate an increased spatial extension, characteristic of halo structures where the valence nucleon has a high probability of existing at large radii, which effectively increases the interaction radius $R_{\rm eff}$. 
Points below the bands often signify that the resonance is not a pure single-particle state, which can arise due to configuration mixing where resonance strength is fragmented, or from spectroscopic factors ($S < 1$) that scale down the observed width. 
Such deviations may also result from the coupling of these resonance states to other open reaction channels or a complex configuration mixing within the wave function, which requires a more sophisticated many-body description than the potential model provided here.

\end{document}